\documentclass[doublespacing]{elsart}

\usepackage{graphicx}
\usepackage{amssymb}

\begin{document}

\begin{frontmatter}

\title{Method of Image and Transmission through Semi-infinite Nanowires}
\author[address1]{Chih-Hui Wu} and
\author[address1,address2]{Chung-Yu Mou}
\address[address1]{Department of Physics,
National Tsing Hua University, Hsinchu 30043, Taiwan}

\address[address2]{National Center for Theoretical Sciences,
P.O.Box 2-131, Hsinchu, Taiwan}

\begin{abstract}
The method of functional integral bosonization is extended to
examine the transmission properties of semi-infinite nanowires. In
particular, it is shown that edge states will arise at the end point
of the dimerized semi-infinite spin-chain and by combining the
method of image and the bosonization technique, the system can be
properly bosonized. Based on the bosonized action and a
renormalization group analysis, it is shown that unlike scattering
due to single bulk impurity in the nanowire, the scattering
potential remains relevant even for slightly attractive potential
due to the interaction between the edge state and its image. When
the strength of potential goes beyond a critical strength, the tip
of the semi-infinite nanowire may become insulating.
\end{abstract}

\begin{keyword}
nanowire \sep method of image \sep bosonization
\PACS 74.20.-z \sep 74.80.-g \sep 74.25.Fy \sep 74.50.+r
\end{keyword}

\end{frontmatter}

\section{Introduction}
Nanowires have been considered as ideal tools for probing
nano-materials\cite{Ho1,Ho2}. In real applications, the scanning
probe microscopes (SPM) represent the typical realization\cite{SPM}.
In this approach, the tip of the probe is crucial. It is therefore
important to understand the properties of the tip. From theoretical
point of view, the tip can be ideally considered as the end of a one
dimensional nanowire. The physics is thus embedded in the
semi-infinite nanowires. In various situations, it is known that
edge states may arise at the end. Nevertheless, conventional studies
of the 1D wire have mostly been focused on its bulk properties,
whereas assembled nanowires can only have finite lengths and must
terminate at some sites (the ends, or the edges). It is therefore
important to investigate effects that are due to the ends.

In the absence of interactions, the presence of the edge can be
handled by method of image or its
generalization\cite{mou1,mou11,mou12}. In this case, effects due to
the ends are identified and reflected in the tunneling spectrum. For
example, because the end couples $k$ and $-k$, the wavefunction is
modulated by $sin(kx)$ so that the van Hove singularity is
suppressed in the local density of states near the
end\cite{mou1,mou11,mou12}. The more interesting situation occurs
when the reflection symmetry is broken. In that case, it is shown
that localized edge states may arise and result in peaks in the
tunneling spectrum\cite{mou2}. These conclusions are based on free
electrons in which the problem is linear so that method of image is
applicable. In real materials, particularly in nano-materials,
interactions can not be neglected. In that case, the applicability
of method is called into question. It is therefore of important to
examine the applicability of method of image or find a general
method to handle the boundary.

In this work, we shall show that indeed the method of image can be
generalized even in the presence of interactions. The generalization
results from a combination of method of image and the bosonization
technique. By adopting the method of the functional bosonization,
recently developed by Yurkevich\cite{ref1}, we show that the
boundary conditions can be easily implemented. The marriage of the
two methods allows one to investigate effects due to the edge states
in the most general situations. To illustrate our method, we shall
consider the simplest situation that supports the edge states :
spinless electrons in dimerized semi-infinite nanowires, which would
correspond to fermionized spin-chains. We shall first establish the
existence of the edge state in the presence of interactions. This
will be investigated by considering a dimerized spin chain. The
method of functional bosonization is then combined with the method
of image to calculate the two-point correlation function for
dimerized nanowires. Finally, the bosonization of the partition
function for semi-infinite nanowire will be derived. Based on the
derived partition function, we perform a renormalization group (RG)
analysis and obtain the RG flow equation. The implication of the
effect of the edge state on itinerary electrons will be discussed.

\section{Existence of the edge state}
We shall first illustrate that under appropriate conditions,
the edge states exist and persist even in the presence
of strong interaction. For this purpose, we
start by considering a dimerized semi-infinite spin-chain in which the
transverse spin coupling changes alternatively. The Hamiltonian can be
written as
\begin{equation}
H_{S}=\sum\limits_{i=1}^{i= \infty
}2J_{i}(S_{i}^{x}S_{i+1}^{x}+S_{i}^{y}S_{i+1}^{y})
+J_{z}(S_{i}^{z}+1/2)(S_{i+1}^{z}+1/2)+hS_{i}^{z}, \label{H_S}
\end{equation}
where $h$ is the magnetic field, $J_{i}=t_{1}$ when $i$ is odd and $%
J_{i}=t_{2}$ when $i$ is even. By applying the Jordan-Wigner transformation $%
\psi _{i}^{\dagger }=K(i)S_{i}^{+}$ and $\psi _{i}=S_{i}^{-}K(i)$ with $%
K(i)=exp[i\pi \sum_{j}^{i-1}(S_{j}^{z}+1/2)]$, $H_{S}$ is then
mapped into the Hamiltonian of spinless electrons with interaction
between nearest neighbors
\[
H=\sum_{i=1}^{i=\infty }J_{i}(\psi _{i}^{\dagger }\psi _{i+1}+\psi
_{i+1}^{\dagger }\psi _{i})+J_{z}\psi _{i}^{\dagger }\psi _{i}\psi
_{i+1}^{\dagger }\psi _{i+1}+h(\psi _{i}^{\dagger }\psi _{i}-1/2).
\]
Here the applied $h$ field is equivalent to the chemical potential.
In the absence of the $J_{z}$ term, $H$ reduces to free electron
model with dimerized hopping amplitudes. This is the model that can
describe polyacetylene\cite {Su1}. In this case, the bulk dispersion
can be solved exactly and is given by $\varepsilon
_{k}^{2}=t_{1}^{2}+t_{2}^{2}+2t_{1}t_{2}cos2ka$ with $a$ being the
lattice constant. It is clear that the spectrum has a gap $2\Delta
=2|t_{2}-t_{1}|$. For semi-infinite chains, previous studies show
that when $t_1 < t_2$, there is an edge state that decays
exponentially from the edge with energy right at zero\cite{mou2}. It
is conceivable by continuity that the edge state survives when $J_z$
is turned on as long as $J_z$ is small in comparison to the gap. To
check how the edge state evolves in the presence of interaction, we
solve $H_S$ (instead of $H$) by resorting to method of exact
diagonalization. The energy of the edge state can be found by tuning
$h$ and measuring the local density of electrons at the 1st site. As
$h$ sweeps across the edge state, a large change of the local
density occurs due to the fact that the edge state is localized. In
Fig.~1, we show the numerical result of 7 sites of spin chain. The
jump in the local density is clearly manifested.

The energy of the edge state can be read out by finding the
magnitude of $h$ at the jump. In Fig.~2, we show the energy of the
edge state versus $J_z$. Clearly, one sees that in presence of
interactions, the edge state persists except that the energy is no
longer right at zero.

\section{Two-point Green's functions in dimerized infinite chains}

To understand how the localized edge state interacts with extended
states, we shall first consider the case without the boundary to
illustrate the effect of dimerization on the extended states. In
this case, the Hamiltonian of the extended states for spinless
electrons is
\begin{eqnarray}
H= \sum t_{i} \psi_{i}^{\dagger} \psi_{i+1}+h.c.+
V n_{i} n_{i+1}.
\end{eqnarray}
Here when $i$ is odd, $t_{i}=t_{1}$, otherwise, $t_{i}=t_{2}$.
In the absence of
$V$, there are two bands separating by a gap. We shall set
the Fermi energy $\varepsilon_{f}$ in the upper band.
After the Fourier transformation, the kinetic term becomes
\begin{eqnarray} \label{3.3}
H_0= \sum_k (\psi_{odd}^{\dagger}(k),\psi_{even}^{\dagger}(k)) \left(
\begin{array}{cc}
0 & \Delta_k \\
\Delta^{\ast}_k & 0
\end{array}
\right) \left(
\begin{array}{l}
\psi_{odd}(k) \\
\psi_{even}(k)
\end{array}
\right) \quad,
\end{eqnarray}
where $\Delta_k=t_{1}e^{i k a}+t_{2}e^{-i k a}$.
Note that because the
lattice consists of two sublattices: even sites and odd sites,
the theory is a two-component theory with the
obvious notations: $\psi_{odd}$ represents $\psi$ at odd sites
and $\psi_{even}$ represents $\psi$ at even sites.
To focus on the itinerant
electrons, we linearize $H_0$ in k near the Fermi wavevector $\pm k_{f}$.
By replacing $i(k \pm k_f)$ by $\partial_x$,
the action for the kinetic terms becomes
\begin{eqnarray}
S_0 &=& \int dx d \tau \hat{\psi_L}^{\dagger}\left(
\begin{array}{cc}
\partial_{\tau}-\varepsilon_f & \varepsilon_{f}+i\tilde{v_f}\partial_{x} \\
\varepsilon_{f}+i\tilde{v_f}^{\ast}\partial_{x} & \partial{\tau-\varepsilon_f
}
\end{array}
\right) \hat{\psi_L} \\ \nonumber
& & + \hat{\psi_R}^{\dagger}\left(
\begin{array}{cc}
\partial_{\tau}-\varepsilon_f & \varepsilon_{f}-i\tilde{v_f}^{\ast}\partial_{x}
\\
\varepsilon_{f}-i\tilde{v_f}\partial_{x} & \partial{\tau-\varepsilon_f
}
\end{array}
\right) \hat{\psi_R}. \label{kinetic}
\end{eqnarray}
Here $\tilde{v_f}=a(2t_1t_2sin2k_fa-it_1^2+it_2^2)/\varepsilon_f$ and
phase factors, $
e^{i\delta}=(t_1e^{ik_fa}+t_2e^{-ik_fa})/\varepsilon_f$,
have been absorbed into $\psi_{even}^L$ or $\psi_{even}^R$.
$R$ and $L$ represent right and left movers with corresponding
energy dispersions: $i \omega = - v_f k$ and $i \omega = v_f k$,
found by diagonalizing $S_0$\cite{uniform}.
Note that $v_f=Re[\tilde{v_f}]$ is the Fermi velocity.

In the absence of interaction,
the Green's function can be found
by rotating to a new basis so that the action is diagonalized.
For example, for
the left mover, the relation between the new basis
and the old basis is
$(c_{upper}^L,c_{lower}^L)^T\equiv \hat{M}(\psi_o^L,\psi_e^L)^T$
with $\hat{M}$ being given
by (in the $k$ space)
\begin{eqnarray}  \label{3.15}
\hat{M}=\frac{1}{\sqrt{2}}\left(
\begin{array}{cc}
exp[\frac{i Im[\tilde{v_f}]k}{2\varepsilon_f}] & exp[\frac{-i Im[\tilde{v_f}%
]k}{2\varepsilon_f}] \\
exp[\frac{i Im[\tilde{v_f}]k}{2\varepsilon_f}] & -exp[\frac{-i Im[\tilde{v_f}%
]k}{2\varepsilon_f}]
\end{array}
\right).
\end{eqnarray}
To the 1st order in $k$, the action for the left mover becomes
\begin{eqnarray}
S_0^L \approx \sum_\omega \int dk \hat{c}^{L\dagger}\left(
\begin{array}{cc}
i \omega -v_f k & 0 \\
0 & i\omega-2\varepsilon_f+v_fk
\end{array}
\right) \hat{c}^L,
\end{eqnarray}
where $\hat{c}^L=(c^L_{upper},c^L_{lower})^T$. Obviously, $c_{upper}^L$ and $
c_{lower}^L$ correspond to bonding and
anti-bonding bands respectively. Since the lower band is filled and so we
are left with only $c_{upper}^L$ whose two-point correlation function is
\begin{eqnarray}  \label{9}
<c_{upper}^{\dagger}c_{upper}>=\frac{1}{\beta v_f sin[\frac{\pi}{\beta}%
(i(x-x^{\prime})/v_f+(\tau-\tau^{\prime}))]}.
\end{eqnarray}
By using Eq.(\ref{3.15}), the relations between $c$ and $\psi$ are
\begin{eqnarray}  \label{3.18}
\big\{
\begin{array}{l}
\psi_o^R(x,\tau)=\frac{1}{\sqrt{2}}c_{upper}^L(x-\frac{Im[\tilde{v_f}]}{%
2\varepsilon_f},\tau) \\
\psi_e^R(x,\tau)=\frac{1}{\sqrt{2}}c_{upper}^L(x+\frac{Im[\tilde{v_f}]}{%
2\varepsilon_f},\tau)
\end{array}
\quad.
\end{eqnarray}
We thus obtain the Green's functions for electrons
\begin{eqnarray}  \label{3.19}
\begin{array}{l}
G_{o\rightarrow o}^L=G_{e\rightarrow e}^L=\frac{1}{2\beta v_f
sin[\pi((\tau-\tau^{\prime})+i(x-x^{\prime})/v_f)/\beta]}, \\
G_{o\rightarrow e}^L=\frac{1}{2\beta v_f
sin[\pi((\tau-\tau^{\prime})+i(x-x^{\prime}-\frac{Im[\tilde{v_f}]}{
\varepsilon_f})/v_f)/\beta]}, \\
G_{e\rightarrow o}^L=\frac{1}{2\beta v_f
sin[\pi((\tau-\tau^{\prime})+i(x-x^{\prime}+\frac{Im[\tilde{v_f}]}{
\varepsilon_f})/v_f)/\beta]}.
\end{array}
\end{eqnarray}
Here except for the position shift,
$\Delta \equiv \frac{Im[\tilde{v_f}]}{
\varepsilon_f}$,
 between even sites and odd sites,
the Green's functions are the same as those for a single band. From
the point of view of the wavefunction, if one denotes $\psi_{odd} =
e^{ikx}$, one can write $\psi_{even} = e^{ikx - i \delta (k)}$ with
$\delta (k)$ being the phase shift. Thus linearizing with respect to
$k_f$ results in a shift of $x$. We then obtain
\begin{equation}
\Delta = \partial_k \delta (k_f) = -a \frac{t^2_2 - t^2_1}
{t^2_1 + t^2_2 +2t_1t_2 \cos 2 k_f}, \label{phaseshift}
\end{equation}
Therefore, the position shift can be attributed to the phase shift
between even sites and odd sites. Clearly, when $t_2\gg t_1$,
$\Delta$ approaches $-a$. In this case, $ G_{odd\rightarrow
even}^L(x-x^{\prime})=G_{odd\rightarrow odd}^L(x-(x^{\prime}+a))$.
This is consistent with the fact that when $t_2\gg t_1$, electrons
on two neighboring sublattice points have the same amplitudes.

We now include the effect of the interaction which
takes the following form in the continuum approximation
\begin{equation}
\frac{1}{4}\hat{V}_0=\int dxdx^{\prime}\frac{1}{4}(\rho_o(x),\rho_e(x)) \left(
\begin{array}{cc}
0 & V_0 \\
V_0 & 0
\end{array}
\right) \left(
\begin{array}{l}
\rho_{o}(x^{\prime}) \\
\rho_e(x^{\prime})
\end{array}
\right),
\end{equation}
where $\rho=\rho^R+\rho^L$ is the density operator at odd sites($o$) or even
sites($e$)
and $V_0(x,x^{\prime}) = \frac{1}{2}V_0
[\delta(x-x^{\prime}-a)+\delta(x-x^{\prime}+a)]$.
Following \cite{ref1}, we first apply the Hubbard-Stratonovinch (HS)
transformation
to decouple the density operators
\begin{equation}  \label{3.21}
e^{-\frac{1}{4}\hat{V}_0} = \frac{1}{Z} \int  D \sigma
e^{-\hat{V}_0^{-1}-i(\sigma_o\rho_o+%
\sigma_e\rho_e)},
\end{equation}
where $\hat{V}_0^{-1}=(\sigma_o,\sigma_e) \left(
\begin{array}{cc}
0 & V_0^{-1} \\
V_0^{-1} & 0
\end{array}
\right) \left(
\begin{array}{l}
\sigma_{o} \\
\sigma_e
\end{array}
\right)$ and $Z=\int D
\sigma e^{-\hat{V}_0^{-1}}$. For fixed $\sigma$, the $-i\sigma \rho$
term in combination with Eq.(\ref{kinetic}) constitute the quadratic
term.
However, due to the matrix
nature of the action, the $\sigma$ fields can not be expressed as a
phase factor,$e^{i\phi}$, as what happens in the case when $t_1=t_2$\cite{ref1}.
To avoid this difficulty, we first perform the transformation
of Eq.(\ref{3.15}). We shall take the left mover
as a demonstration. The right mover can be handled by
the same method.  To 1st order in $k$, the
transformed action $\hat{M}S _\psi^L\hat{M}^{-1}$ becomes
\begin{eqnarray}  \label{3.23}
S_{c}^L= \int d\tau \int dx \hat{c}^{L\dagger}\left(
\begin{array}{cc}
\partial_{\tau}+i v_f\partial_x-i\sigma_+ & -i\sigma_- \\
-i\sigma_- & \partial_{\tau}-2\varepsilon_f-iv_f\partial_x-i\sigma_+
\end{array}
\right)\hat{c}^L,
\end{eqnarray}
where $\sigma_+=(\sigma_o+\sigma_-)/2$ and $\sigma_-=(\sigma_o-\sigma_e)/2$.
In the above derivation, we have neglected the commutator
of $\partial_{x}$ and $
\sigma$ because it is of 2nd order in $V_0$ and $1/\beta$. Furthermore,
its effect
is to induce a shift in the position of $\sigma$; equivalently, it induces
a shift in the space dependence of $V_0(x,x^{\prime})$.
Since we shall be concerning the long distance behavior, this position shift
should not matter.
Because the anti-bonding states are $2\varepsilon_f$ below the Fermi energy,
we can drop the anti-bonding states and keep the bonding states with
the action being given by
\begin{equation}  \label{3.24}
S_L= \int d\tau \int dx c_{upper}^{L\dagger}(x^{\prime},\tau)(\partial_{\tau}+i
v_f\partial_x-i\sigma_+)c_{upper}^L(x,\tau)\quad.
\end{equation}
At this stage, the effect of $\sigma$ can be absorbed as a phase factor
by defining
\begin{equation}  \label{3.25}
c_{upper}^L(x,\tau)\equiv\tilde{c}_{upper}^L(x,\tau)e^{i\phi_L(x,\tau)}
\quad.
\end{equation}
Here if we require
\begin{equation}  \label{3.26}
(\partial_{\tau}+iv_f\partial_x)\phi_L(x,\tau)=\sigma_+.
\end{equation}
the $i\sigma \rho$ term is canceled out so that
$\tilde{c}_{upper}^L$ is completely free and its two-point
correlation function of $\tilde{c}_{upper}^L$ is exactly the same as
Eq.(\ref{9}). As a result, the correlation function
$<c_{upper}^Lc_{upper}^{L\dagger}>$
is determined by $<e^{i\phi^L(x,\tau)-i%
\phi^L(0,0)}>$.
Furthermore, the relation between $\sigma_+$ and $\phi_L$
in Eq.(\ref{3.26}) implies that
to evaluate $<e^{i\phi^L(x,\tau)-i\phi^L(0,0)}>$, one needs to find
the Lagrangian for $\sigma_+$.

The Lagrangian for $\sigma_+$ can be found by
first noting that the transformation of
Eq. (\ref{3.25}) induces a Jacobian
\begin{eqnarray}
lnJ^L=Tr[ln \frac{\left(
\begin{array}{cc}
\partial_{\tau}-\varepsilon_f-i\sigma_o & \varepsilon_{f}+i\tilde{v_f}%
\partial_{x} \\
\varepsilon_{f}+i\tilde{v_f}^{\ast}\partial_{x} & \partial_{\tau}-%
\varepsilon_f-i\sigma_e
\end{array}
\right)}{\left(
\begin{array}{cc}
\partial_{\tau}-\varepsilon_f & \varepsilon_{f}+i\tilde{v_f}\partial_{x} \\
\varepsilon_{f}+i\tilde{v_f}^{\ast}\partial_{x} & \partial_{\tau}-%
\varepsilon_f
\end{array}
\right)}]
\end{eqnarray}
By expanding in powers of $\sigma$, the above Jacobian becomes
\begin{eqnarray}
lnJ^{L/R}=-\sum_{n=1}^{\infty}\frac{1}{n}Tr( i \hat{ \sigma} \hat{G}^{L/R}
)^n\quad.
\end{eqnarray}
where $\hat{G}$ is  the $2\times 2$ matrix of the Green's function
in the absence of interactions and
\begin{eqnarray}
\hat{\sigma} =
\left(
\begin{array}{cc}
\sigma_o & 0 \\
0 & \sigma_e
\end{array}
\right).
\end{eqnarray}
Using the loop cancelation theorem\cite{ref1}, we retain only the
$n=2$ term
\begin{eqnarray}
\frac{1}{2}(\sigma_o,\sigma_e) \hat{J}^{L/R}\left(
\begin{array}{l}
\sigma_o \\
\sigma_e
\end{array}
\right)=-\frac{1}{2}Tr[\hat{\sigma}\hat{G}^{L/R} \hat{\sigma}
\hat{G}^{L/R}]\quad,
\end{eqnarray}
After adding contributions from left and right movers
and retaining only the lowest orders in $q$ and $\omega$, we obtain
\begin{eqnarray}
\hat{J} \equiv
\hat{J}^{L} + \hat{J}^{R}
\approx\frac{v_f q^2}{2\pi^2\beta (\omega^2+v_f^2 q^2)}\left(
\begin{array}{cc}
1 & 1 \\
1 & 1
\end{array}
\right)\quad .
\end{eqnarray}
Combining $\frac{1}{2}\hat{J}$ with $V_0^{-1}$, we obtain
the action
for the $\sigma$ fields
\begin{eqnarray}
\hat{S}_{\sigma} = \frac{1}{4\pi\beta}
\sum_{\Omega}\int_0^{\infty}
dq
\left[ \sigma_+^{\ast}(V_0^{-1}+\frac{v_f q^2}{%
\pi(\omega^2+v_f^2 q^2)})\sigma_+(q,\omega)
-\sigma_-(q,\omega)^{\ast}V_0^{-1}\sigma_-(q,\omega)
\right]. \nonumber \\
\label{sigma}
\end{eqnarray}
Using Eqs.(\ref{3.26}) and (\ref{sigma}), we obtain
\begin{eqnarray}
&<&\phi_L(x,\tau)\phi_L(0,0)>=lnsin[\frac{\pi}{\beta v_f}(ix+v_f\tau)]
\nonumber \\
&-&\frac{g+g^{-1}}{2}ln|sin[\frac{\pi}{\beta v}(ix+v\tau)]| +iarg[sin[\frac{%
\pi}{\beta v}(ix+v\tau)]], \qquad
\end{eqnarray}
where $ v^2 \equiv  v^2_f (1+V_0 /\pi v_f)$ and $g \equiv v_f/v$.
Therefore, the two-point correlation function of $c_{upper}^L$ is given by
\begin{eqnarray}
& & <c_{upper}^L(x,\tau)c_{upper}^{L\dagger}(x^{\prime},\tau^{\prime})>
\nonumber \\
& & =
\frac{1}{\beta v_f}\frac{e^{i arg[sin[\frac{\pi}{\beta v}%
(i(x-x^{\prime})+v(\tau-\tau^{\prime}))]]}}{ (sin[\pi((\tau-\tau^{%
\prime})+i(x-x^{\prime})/v)/\beta])^{-(g+g^{-1})/2}}.\label{cupper}
\end{eqnarray}
It is clear that the
interaction changes not only the exponent but
also the phase in
correlation functions. Similar procedure applies to
the right mover, the
correlation function for the right mover
turns out to be the complex
conjugate of $<c_{upper}^Lc_{upper}^{L\dagger}>$.
Finally, the two-point Green's functions for electrons are obtained
by combining Eqs. (\ref{3.18}) and (\ref{cupper}). The net effect
of the relation Eq. (\ref{3.18})
is to shift positions of even sites by $\Delta$, given
by Eq.(\ref{phaseshift}).
Clearly, the shift is not changed by
the interaction and the same conclusion applies to
other Green's functions.

We conclude this section by noting that the bosonization of
dimerized chains is almost the same as that for one band model
except when calculating Green's functions, positions have to be
shifted appropriately as indicating in Eq.(\ref{3.19}). Obviously,
this also applies to the general case when the system contains more
than two sublattices. In that case, one first bosonizes the energy
band cut by the Fermi energy. The important information one needs
for calculating Green's functions is then contained in the
transformation matrix, similar to Eq.(\ref{3.15}), which determines
the position shift of each site.

\section{Generalized method of image for semi-infinite chains}

We now apply method of functional bosonization to semi-infinite
chains. A generalized method of image will be developed in the
presence of interactions. We first set the boundary right at site 0,
thus $\psi (0) =0$. In the continuum approximation, the Hamiltonian
for the energy band cut by the Fermi energy can be generally written
as $H=\hat{H}_R+\hat{H}_L+\hat{V}_0$ with the interaction term being
given by
\begin{equation}
\hat{V}_0=\frac{1}{2} \int^{\infty}_0 dx
\int^{\infty}_0 dx^{\prime} \rho(x) V_0(x - x^{\prime})
\rho (x^{\prime}), \label{potential}
\end{equation}
and the kinetic terms be given by
\begin{eqnarray}
\hat{H}_{R/L}=\mp i v_f\int_{0}^{\infty}dx
\psi_{R/L}^{\dagger}\partial_x\psi_{R/L}.
\end{eqnarray}
Here the operator $\psi
\equiv e^{ik_f x}\psi_{R}+e^{-ik_f
x}\psi_{L}$ is the continuum operator for
electrons in the
energy band cut by the Fermi energy. In an infinite chain,
the right-mover and the left-mover are independent.
This independence, however, is lifted in the semi-infinite
chain due to the boundary condition
\begin{equation}
\psi(0)=\psi_R(0)+\psi_L(0)=0\quad.
\end{equation}
To satisfy the boundary condition, it is useful to extend the
defining domain of $\psi_R$ and $\psi_L$ to $x<0$.
For this purpose, we define
\begin{equation}
\psi_{R/L}(-x,\tau)=-\psi_{L/R}(x,\tau), \label{boundary}
\end{equation}
so that the boundary condition is automatically satisfied.
The kinetic term can be then rewritten in terms
of a single chiral field
\begin{eqnarray}
\hat{H}_R+\hat{H}_L=-iv_f\int_{-\infty}^{\infty}dx\psi_{R}^{\dagger}
\partial_x\psi_{R}. \label{right}
\end{eqnarray}
Thus in the absence of interaction,
the Green's function for
the right mover is exactly the same as that of the
infinite chain. By using Eq.(\ref{boundary}),
we find that the full Green's function can be written as
\begin{eqnarray}
g(x,x^{\prime};\tau,\tau^{\prime})=
e^{i k_f (x-x^{\prime})} G^R(x,x^{\prime};\tau,\tau^{
\prime})
+e^{-i k_f (x-x^{\prime})} G^R(-x,-x^{\prime};\tau,\tau^{
\prime}) \nonumber \\
-e^{i k_f (x+x^{\prime})} G^R(x,-x^{\prime};\tau,\tau^{
\prime}) +
-e^{-i k_f (x+x^{\prime})} G^R(-x,x^{\prime};\tau,\tau^{
\prime}), \label{image1}
\end{eqnarray}
where we have denoted
the Green's functions for the right/left mover by
$G^{R/L}$.
Note that Eq.(\ref{boundary}) also implies $ G^R(x,x^{\prime}) =
 G^L(-x,-x^{\prime})$. Therefore,
Eq.(\ref{image1}) can be cast in the form
\begin{equation}
g(x,x^{\prime};\tau,\tau^{\prime})=G(x,x^{\prime};\tau,\tau^{
\prime})-G(x,-x^{\prime};\tau,\tau^{\prime}), \label{image2}
\end{equation}
where
$G(x,x^{\prime};\tau,\tau^{\prime})=
e^{i k_f (x-x^{\prime})} G^R(x,x^{\prime};\tau,\tau^{
\prime})
+e^{-i k_f (x-x^{\prime})} G^L(x,x^{\prime};\tau,\tau^{
\prime})
 $
has the form of the Green's function for the infinite chain. Notice
that without interactions, $G^{R/L}$ are the same as the bulk
Green's functions for infinite chains, thus Eq.(\ref{image2})
demonstrates the validity of method of image.  In the presence of
the interaction, $G^{R/L}$ and $G$ appear not to be the bulk Green's
functions and their meaning requires further exploration (see
below). Nonetheless, the form of Eq.(\ref{image2}) is still valid
except that one has to resort to Eq.(\ref{image1}) to calculate the
Green's function. Obviously, all the information needed is in $G^R$.

To calculate $G^R$, we first apply the HS transformation\cite{ref1} on
the interaction term for $0< x < \infty $
\begin{eqnarray}
e^{-\int^{\beta}_0 d\tau \hat{V}_0 (\tau)} = \int D \sigma e^{-\frac{1}{2}
\int^{\beta}_0 d\tau  \left[ \int_0^{\infty} dx \int_0^{\infty} dx^{\prime}
\sigma V_0^{-1} (\tau) \sigma+i\int_0^{\infty}dx\sigma
c^{\dagger} c \right]} /Z, \label{interaction}
\end{eqnarray}
where $V_0^{-1}(x^{\prime},x^{\prime\prime})$ is the interaction
potential that satisfies $\int
dx^{\prime}V_0(x,x^{\prime})V_0^{-1}(x^{\prime},x^{\prime\prime})=%
\delta(x-x^{\prime\prime})$. Since we shall be interested in
the local approximation in which $x \approx x^{\prime}$ dominates,
by extending the HS field, $\sigma (x, \tau)$,
into $- \infty < x < 0$
\begin{equation}
\sigma(-x, \tau)\equiv \sigma(x, \tau), \label{sigmaeven}
\end{equation}
and assuming that $V_0^{-1}$ is
symmetric in the real space, the first term on the RHS
in Eq.(\ref{interaction}) can be rewritten as
$\frac{1}{4} \int^{\beta}_0 d\tau
\int_{-\infty}^{\infty}dx \int_{-\infty}^{\infty}
dx^{\prime}\sigma(x)\hat{V}
_0^{-1}\sigma(x^{\prime})$.
For the second term on the RHS
in Eq.(\ref{interaction}), we use Eq.(\ref{boundary})
and neglect terms that contain the fast-oscillating factor,
$e^{\pm 2i k_fx}$.
The
action for the right mover then becomes
\begin{eqnarray}
\hat{S}_R &=&\int_0^{\beta}d\tau[\int_{-\infty}^{\infty}dx\psi^{\dagger}_R(
\partial_{\tau}-i\sigma(x)-iv_f \partial_x)\psi_R  \nonumber \\
&+&\frac{1}{4} \int_{-\infty}^{\infty}dx \int_{-\infty}^{\infty}
dx^{\prime}\sigma(x)\hat{V}
_0^{-1}\sigma(x^{\prime})]. \label{S_R}
\end{eqnarray}
We have thus transformed the semi-infinite problem into an infinite
one except that the $\sigma$ field has to be even:
$\sigma(x)=\sigma(-x)$. Clearly, the loop cancelation
theorem\cite{ref1} is still valid so that the action of $\sigma$ is
quadratic. By making use of $\sigma(q,\Omega)=\sigma(-q,\Omega)$, we
obtain
\begin{eqnarray}  \label{4.20}
\hat{S}_{\sigma}=\frac{1}{4 \pi\beta V_0 }\sum_{\Omega}\int_0^{\infty}
dq\sigma(q,\Omega)\frac{\Omega^2+v^2q^2}{\Omega^2+v_f^2q^2}%
\sigma(q,-\Omega),\quad
\end{eqnarray}
where $v^2=v_f^2+\frac{v_f V_0}{\pi}$. Note that
the correlation function of $\sigma$ is given by
\begin{eqnarray}  \label{4.21}
<\sigma(q,\Omega)\sigma(q^{\prime},\Omega^{\prime})>_\sigma=
2\pi\beta V_0\frac{\Omega^2+v_f^2q^2}{\Omega^2+v^2q^2}(\delta(q+q^{
\prime})+\delta(q-q^{\prime}))\delta_{\Omega+\Omega^{\prime}}.
\end{eqnarray}
Here the term $\delta(q+q^{\prime})$,
is due to the
relation $\sigma(q,\Omega)=\sigma(-q,\Omega)$; clearly,
it is
the crucial difference between the infinite chains
and semi-infinite chains and reflects the effects of
boundary condition.

Similar to Eq.(\ref{3.26}), $\sigma$ can be absorbed by a phase
$\phi_R$ with the relation $\phi_R (q,\Omega) = \sigma(q,\Omega) /(i
\Omega +v_f q)$. Therefore, the correlation for the phase can be
computed by using Eq.(\ref{4.20}). After performing appropriate
contour integrations, we find that the analytic form for the
correlation function of the phase is given by
\begin{eqnarray}
& & \langle \phi_R (x,\tau) \phi_R (x^{\prime},\tau^{\prime})
\rangle \nonumber \\
&=& \ln \left[ \sin \{
\frac{\pi}{\beta v_f}
[i(x-x^{\prime})-v(\tau-\tau^{\prime})] \} \right]
-\frac{g+g^{-1}}{2}
\ln \left| \sin \{
\frac{\pi}{\beta v}
[i(x-x^{\prime})-v(\tau-\tau^{\prime})] \} \right| \nonumber \\
&-& i \arg \left[ \sin \{
\frac{\pi}{\beta v}
[i(x-x^{\prime})-v(\tau-\tau^{\prime})] \} \right]
-\frac{g^{-1}-g}{2}
\ln \left| \sin \{
\frac{\pi}{\beta v}
[i(x+x^{\prime})-v(\tau-\tau^{\prime})] \} \right| . \nonumber \\
\label{phi_R}
\end{eqnarray}
The Green's function
of the right mover is obtained by
computing  $e^{-\frac{1}{2}<(\phi_R(x,\tau)-
\phi_R(x^{\prime},\tau^{\prime}))^2>}$, we find
\begin{eqnarray}
G^R (x,x^{\prime};\tau,\tau^{\prime}) &=& \frac{\pi}{v_f}
e^{-iarg[F(x,x^{\prime};\tau,\tau^{\prime})]}
|F(x,x^{\prime};\tau,\tau^{\prime})|^{-\frac{g+g^{-1}}{2}%
}|F(x,-x^{\prime};\tau,\tau^{\prime})|^{-\frac{g-g^{-1}}{2}}  \nonumber \\
&\times&|F(x,-x;\tau,\tau)|^{\frac{g-g^{-1}}{4}}|F(x^{\prime},-x^{\prime};%
\tau^{\prime},\tau^{\prime})|^{\frac{g-g^{-1}}{4}},
\nonumber \\
\label{GR}
\end{eqnarray}
where $F(x,x^{\prime};\tau,\tau^{\prime})=sin[\frac{\pi}{\beta v}
(i(x-x^{\prime})-v(\tau-\tau^{\prime}))]$. From $G^R$, one
obtains the full Green's function by using
Eq.(\ref{image1}) or Eq.(\ref{image2}). Finally, the Green's functions
of real electrons
are obtained by shifting positions according to Eqs.(\ref{3.18})
and (\ref{phaseshift}).

In conclusion, the method
of image, Eq.(\ref{image2}),
is valid even in the presence of interaction except that now
the bulk Green's functions have to be re-interpretated as
the Green's function of the infinite chains with the
even $\sigma$ field.

\section{The effect of the edge state on itinerary electrons}

As demonstrated in Sec. (2), when $t_1 < t_2$, a localized edge
state arises. In the absence of interactions, its energy is right at
zero with the wavefunction be given by $\psi_0 (x)
\approx(1,0,-\epsilon ,0,\epsilon ^{2},0,-\epsilon ^{3},0,\cdot
\cdot \cdot )$ with $\epsilon \equiv
t_{1}/t_{2}$\cite{mou1,mou11,mou12,,mou2}; while in the presence of
interactions, the edge state persists with the energy being shifted
away from zero.

Since the edge state is localized, it causes higher electron
densities near the edge when the Fermi energy exceeds the energy of
the edge state. The density-density interaction then induces the
potential scattering on electrons near the Fermi energy. This is
entirely similar to single bulk impurity in the nanowire, where if
there is no interaction between the impurity and the nanowire, the
level for the electron residing on the impurity also lies outside
the energy bands of the nanowire. The potential scattering due to
single bulk impurity was investigated by Kane and
Fisher\cite{Fisher}. In the following, instead of performing more
detailed analysis such as calculating changes of density of
states\cite{ref2}, we shall follow Kane and Fisher and perform the
stability analysis on the itinerary electrons qualitatively based
the renormalization group analysis. The result will be conceptually
useful in considering the transport properties of semi-infinite
nanowires.

We first note that the electron operator can be represented by $
\hat{\psi}(x) + \hat{\psi}_0(x)$ with $\hat{\psi}(x)$ denote the
operator for itinerary electrons near the Fermi surface.
Substituting this representation into the interaction potential,
Eq.(\ref{potential}), a potential scattering term is induced:
\begin{equation}
\hat{V}_p = \int^{\infty}_0 dx \int^{\infty}_0 dx^{\prime}
V_0(x,x^{\prime}) \langle
\psi_0(x^{\prime})^{\dagger}\psi_0(x^{\prime}) \rangle
\hat{\psi}^{\dagger}(x)\hat{\psi} (x). \end{equation}
In the
continuum approximation, we have $ \hat{\psi}(x) \equiv e^{ik_fx}
\psi_R(x)+e^{-ik_fx}\psi_L(x)$ so that
\begin{eqnarray}
\hat{V}_p &= & v_f \int^{\infty}_0 dx v(x)
[\psi_R^{\dagger}(x)\psi_R(x) + \psi_L^{\dagger}(x)\psi_L(x) ]
\nonumber \\
&+&
\int^{\infty}_0 dx
[\lambda(x)\psi_R^{\dagger}(x)\psi_L(x) +
\lambda(x)^{\ast}\psi_L^{\dagger}(x)\psi_R(x) ] ,
\label{semipotential}
\end{eqnarray}
where the second equality is obtained by writing $\hat{\psi}$
in terms of $\psi_R$ and
$\psi_L$
with
$ v_f v (x) \equiv \int^{\infty}_0 dx^{\prime} V_0(x,x^{\prime})
| \psi_0(x^{\prime}) |^2 $ and
$ \lambda (x)  \equiv \int^{\infty}_0 dx^{\prime} V_0(x,x^{\prime})
| \psi_0(x^{\prime}) |^2 e^{i2k_f x} $.
By using Eq.(\ref{boundary}) and defining $v(-x) \equiv v(x)$
and $\lambda (-x) \equiv \lambda^{\ast} (x)$,
the integration domain can be extended to $-\infty$. After being
combined with the kinetic term, Eq.(\ref{right}), the first term
on the RHS of Eq.(\ref{semipotential}) becomes
$ -iv_f \int_{-\infty}^{\infty}dx\psi_{R}^{\dagger}
[\partial_x - i v(x) ] \psi_R (x) $. Therefore, the first term can
be absorbed into the phase of $\psi_R$ by changing
$\psi_R \rightarrow e^{i \int^x dy v(y)}
\psi_R $. The action induced by
the edge state, $S_{edge}$, is thus determined by
the second term.
By writing $\psi = \psi e^{i\phi}$,
$\lambda = |\lambda| e^{i\eta}$,
and using Eq.(\ref{boundary}), we find
\begin{eqnarray}
S_{edge}&=&- \int^{\beta}_0 d\tau \int^{\infty}_{-\infty} dx
|\lambda (x)|
 [e^{i\eta} \psi_R^{\dagger}(x, \tau) \psi_R(-x,\tau)
e^{i\phi_R(-x,\tau)-i\phi_R(x,\tau)}
\nonumber \\
& &
+ e^{-i\eta} \psi_R^{\dagger}(-x, \tau)\psi_R(x,\tau)e^{i\phi_R(x,\tau)-
i\phi_R(-x,\tau)}]. \label{Sedge}
\end{eqnarray}
where we have made use of the relation $\phi_R(x)=\phi_L(-x)$, which
results from the mirror extension of the $\sigma$ field
[see Eq.(\ref{sigmaeven})]
 and the relation between $\sigma$ and the phases $\phi_{R/L}$
[see Eq.(\ref{3.26})].
Obviously, the relevant phase is
$\phi_2(x) \equiv \frac{1}{2i}(\phi_R(x)-\phi_R(-x))$, whose
correlation function can be found by using Eqs.(\ref{4.21}) and (\ref{3.26}).
We obtain
\begin{eqnarray}  \label{5.9}
<\phi_2(\xi)\phi_2(\xi^{\prime})>=\frac{1}{2}ln[\frac{|sin(z-z^{\prime})|^g}{%
|sin(z_f-z_f^{\prime})|}\frac{|sin(z_f-\overline{z_f^{\prime}})|}{|sin(z-%
\overline{z^{\prime}})|^g}],\quad
\end{eqnarray}
where $\xi=(x,\tau)$, $z_f =\frac{\pi}{\beta} (\tau+i x/v_f)$, $z
=\frac{\pi}{\beta} (\tau+i x/v)$, and $\overline{z}$ is the complex
conjugate of $z$. Note that terms that contain the factor
$z_f-\overline{z^{\prime}}_{f}$ or $z -\overline{z^{\prime}}$ are
contributions from the image, representing the boundary effects.
Following Ref.\cite{ref2}, the partition function with the edge
state $Z_{edge} = <e^{-S_{edge}}>$ can be expanded as
\begin{eqnarray}
&&Z_{edge} =\sum_{n,n_1+n_2=n} \frac{1}{n_1!n_2!}
\int \prod_i^n dx_i d\tau_i |\lambda (x_i)|
\nonumber \\
&& \langle \prod_i^{n_1}\psi_R^{\dagger}(\xi_i)\psi_R(\bar{\xi}_i)e^{2\phi_2(
\xi_i) + i\eta}
\prod_{j=n_1+1}^{n_1+n_2}\psi_R^{\dagger}(\bar{\xi}_j)\psi_R(\xi_j)e^{-2
\phi_2(\xi_j) - i \eta} \rangle , \nonumber \\
\label{partition}
\end{eqnarray}
where $\bar{\xi}_i=(-x_i,\tau_i)$. By defining
$\zeta_{f,i}=z_{f,i}$ for $i\leq n_1$
and $\zeta_{f,i}=\overline{z_{f,i}}$ for $i>n_1$ and
applying the Wick's theorem for averaging over $\psi_R$,
we find
\begin{eqnarray}
\langle \prod_i^{n_1}\psi_R^{\dagger}(\xi_i)\psi_R(\bar{\xi}_i)%
\prod_{j=n_1+1}^{n_1+n_2}\psi_R^{\dagger}(\bar{\xi}_j)\psi_R(\xi_j) \rangle
=(\frac{1}{2\beta})^n \det [\frac{1}{sin(\zeta_{f,i}-\overline{\zeta_{f,j}
})}]. \nonumber  \\
\label{psiaverage}
\end{eqnarray}
On the other hand, the average of the phase field $\phi_2$ can be written
as
\begin{eqnarray}  \label{5.13}
<e^{2\sum_i^{n_1}\phi_2(\xi_i)-2\sum_{j=n_1+1}^{n_1+n_2}\phi_2(%
\xi_j)}>=[P(z)]^g/P(z_f),
\end{eqnarray}
where
\begin{eqnarray}
P(z)=\prod_i\frac{1}{|sin(\zeta_i-\overline{\zeta_i})|}\prod_{i\neq j}^n
\frac{|sin(\zeta_i-\zeta_j)|^2}{|sin(\zeta_{i}-\overline{\zeta_{j}})|^2}.
\end{eqnarray}
By using the Cauchy formula\cite{cauchy}, one obtains
$ \det [1/sin(\zeta_{f,i}-\overline{\zeta_{f,j}})]=i^{n_2-n_1}P(z_f)$.
Therefore, $P(z_f)$ in Eq.(\ref{5.13}) gets cancels exactly.
As a result, only the velocity $v$ is retained in the
partition function, while terms that depend on $ix\pm v_f\tau$ are absent.
We obtain
\begin{eqnarray}
Z_{edge} =\sum_{n,n_1+n_2=n}\frac{1}{(2\beta)^n}\frac{1}{n_1!n_2!}%
\int\prod_i^n dx_i d\tau_i |\lambda (x_i)|
\left\{ e^{i (\eta - \pi/2) (n_1-n_2) } [P(z)]^g \right\}. \nonumber \\
\label{partition1}
\end{eqnarray}
We now bosonize this system by showing that $Z_{edge}$
can be reproduced by a bosonic field $\vartheta$.
For this purpose, we first consider a free boson field,
$\varphi (x)$, described by the action
\begin{eqnarray}
S_0 =  \frac{1}{8\pi g}\int dxd\tau \frac{1}{v}
(\partial_\tau\varphi)^2+v(\partial_x\varphi)^2.
\label{Sphi}
\end{eqnarray}
 The correlation
function for $\varphi$ is $-2g \ln | \sin ({z-z^{\prime}}) |$. To
include the effect of boundary, we include the image field $\varphi
(-x)$. The $\vartheta$ is then constructed as
\begin{eqnarray}
\vartheta(x,\tau)=\frac{1}{\sqrt{2}}(\varphi(x,\tau)-\varphi(-x,\tau)).
\label{theta}
\end{eqnarray}
Hence the correlation function for $\vartheta$ is
\begin{eqnarray}  \label{5.18}
\langle \vartheta(x, \tau) \vartheta (x^{\prime}, \tau^{\prime})
\rangle = -2g \left[ \ln | \sin(z-z^{\prime})|- \ln | \sin (z-
\overline{z^{\prime}})| \right] .
\end{eqnarray}
It is then straightforward to verify that $Z_{edge}$
can be written as
\begin{eqnarray}
Z_{edge}= \langle exp[\frac{1}{2\beta}\int^{\beta}_0 d\tau
\int^{\infty}_{-\infty} dx |\lambda (x)|
(e^{i\vartheta + i \eta- i \pi/2}+e^{-i\vartheta - i \eta+ i \pi/2})]
\rangle . \label{Zboson}
\end{eqnarray}
In other words, after the bosonization, the action induced by
the edge state is
\begin{eqnarray}
S_{edge}=-\frac{1}{\beta} \int^{\beta}_0 d\tau \int^{\infty}_{-\infty} dx
|\lambda (x)| \cos [\vartheta (x,\tau)+ \eta-\pi/2]. \label{Sedgeboson}
\end{eqnarray}
Clearly, $S_{edge}$ has the same form as the action
induced by impurities in a bulk Luttinger liquid\cite{ref2}.
However, the phase field $\vartheta$ involved is different.
As shown in Eq.(\ref{theta}), the main difference lies in
the contribution from the image field
$\phi (-x, \tau)$.
Following \cite{Fisher}, we perform a gradient expansion.
First, since in the local approximation, $V(x,x^{\prime})
\approx V_0 \delta (x-x^{\prime})$,
we have $ \lambda (x) \approx V_0 | \psi_0(x) |^2 e^{i2k_f x}$.
For $x>0$, $| \psi_0(x) |^2 \approx e^{-2(x-a)/\xi} $
with $\xi = (\ln t_2/t_1) a$ and $a$ being the lattice constant, $\lambda (x)$
peaks at $x \approx a$. Therefore, after taking $\lambda(-x) =
\lambda(x) ^{\ast}$ into consideration, in the gradient expansion,
$|\lambda (x)|$ can be approximated by $\lambda_0 [\delta (x-a) +
\delta (x+a)]$ with $\lambda_0
\approx | V_0 \int dx |\psi_0|^2 e^{i 2 k_f x} |$.
As a result, the action induced by the edge
reduces to
\begin{eqnarray}
S_{edge}=-\frac{\lambda_0 \sin \eta }{\beta}
\int^{\beta}_0 d\tau
 \cos [\vartheta (a,\tau)]. \label{Sedgeboson1}
\end{eqnarray}
Note that $\eta$ is the phase of $V_0 \int dx |\psi_0|^2 e^{i 2 k_f x}$.
By setting
$k_f \approx 1/a$,
$\eta \approx \tan^{-1} \xi /a = \tan^{-1}
(\ln t_2/t_1) + \pi sign(V_0)$.
Hence when $t_2=t_1$, we obtain $\eta=0$ (repulsive) or $\psi$ (attractive)
and thus
$S_{edge}=0$, which is in consistent
with the fact that the edge state disappears in this case.

To completely describe the edge, one needs to specify
the action of $\vartheta (x,\tau)$
at $x \approx a$. For this purpose, we integrate out $\phi(x)$
in Eq.(\ref{Sphi}) for
all $x$ except for $x=\pm a$ and obtain
\begin{eqnarray}
S_0 = \frac{1}{4\pi g} \sum_{n}
\frac{ | \omega_n | }{ 1- e^{-4| \omega_n/v| a}}
\Phi ^{T}(-i\omega _{n})\left(
\begin{array}{cc}
1 & -e^{-2|\omega _{n}/v|a} \\
-e^{-2|\omega _{n}/v|a} & 1
\end{array}
\right) \Phi (i\omega _{n}). \nonumber \\
\label{Sphi1}
\end{eqnarray}
Here $\omega _{n}$ is the Matsubara frequency and
$\Phi^T (i\omega _{n}) \equiv [
\phi (a,i\omega _{n}),
\phi (-a,i\omega _{n}) ]$. By diagonalizing $S_0$,
we find that the eigenmodes consist of symmetric
and anti-symmetric combination of $\phi(a, i\omega_n)$
and $\phi(-a, i\omega_n)$. Since only the
anti-symmetric mode is involved in $S_{edg}$,
the action for $\vartheta$ is
\begin{eqnarray}
S^{s}_0 = \sum_{n} \frac{1}{4\pi g}
\frac{ | \omega_n | }{ 1- e^{-2| \omega_n/v| a}}
|\vartheta (a,i \omega _{n})|^2.
\label{Stheta2}
\end{eqnarray}
Following \cite{Fisher}, we perform the renormalization-group (RG)
transformation. To leading order, the RG flow equation
is
\begin{eqnarray}
\frac{ d \lambda_{\eta} }{dl} = [1- g (1-e^{-2 \Lambda a/v}) ]\lambda_{\eta},
\label{RG}
\end{eqnarray}
where $\lambda_{\eta} \equiv \lambda_0 \sin \eta$ and $\Lambda$ is a
high-frequency cutoff. Eq.(\ref{RG}) is almost the same as the RG
flow equation for a single impurity in the bulk Luttinger liquid
except for the correction term, $e^{-2 \Lambda a/v}$, due to the
interaction of the edge state with its image. Since $v =v_f /g$ and
$\Lambda \approx O( k_f v_f)$, $e^{-2 \Lambda a/v} = e^{-\alpha g}$
with $\alpha$ being a numerical factor of order $O(1)$. The critical
value of $g$ below which $\lambda_{\eta}$ grows and is relevant is
determined by $1=g_c (1-e^{-\alpha g_c})$. Clearly, the exact
magnitude of $g_c$ depends on the microscopic details through
$\alpha$ and is not universal. Nonetheless, qualitatively for $V_0
> \pi v_f (1/g^2_c -1)$, $V_0$ is relevant and grows indefinitely
under RG transformation.  The tip of the nanowire thus may become
insulating. This effect is entirely due to the interaction between
the edge state and its image. As a result, the critical strength for
the tip of the nanowire being insulating is slightly negative.

\section{Conclusion and acknowledgments}
In conclusion, we have generalized the method of image to be valid
in the presence of interactions. The generalization results from a
combination of method of image and the bosonization technique. The
marriage of the two methods allows one to investigate effects due to
the edge states in the most general situations. Based on the derived
partition function, it is demonstrated that unlike scattering due to
single bulk impurity in the nanowire, the critical strength of
interaction when the tip of the semi-infinite nanowire becomes
insulating is shifted to be slightly attractive due to the
interaction between the edge state and its image. We gratefully
acknowledge discussions with Prof. Hsiu-Hau Lin and  the support
from the National Science Council of the Republic of China.

\newpage
\section*{Figure Captions}

Fig. 1 The local density at the 1st site versus $h$ for zero
temperature and $1/k_BT=100$. Here $t_1=1$, $t_2=4$, and $J_z=0.1$.
\\

Fig. 2 The shift of the edge state energy, i.e., with increasing
$J_z$ in zero temperature.

\newpage
\section*{Figures}

\begin{figure}[\h]
 \includegraphics[width=5.0in]{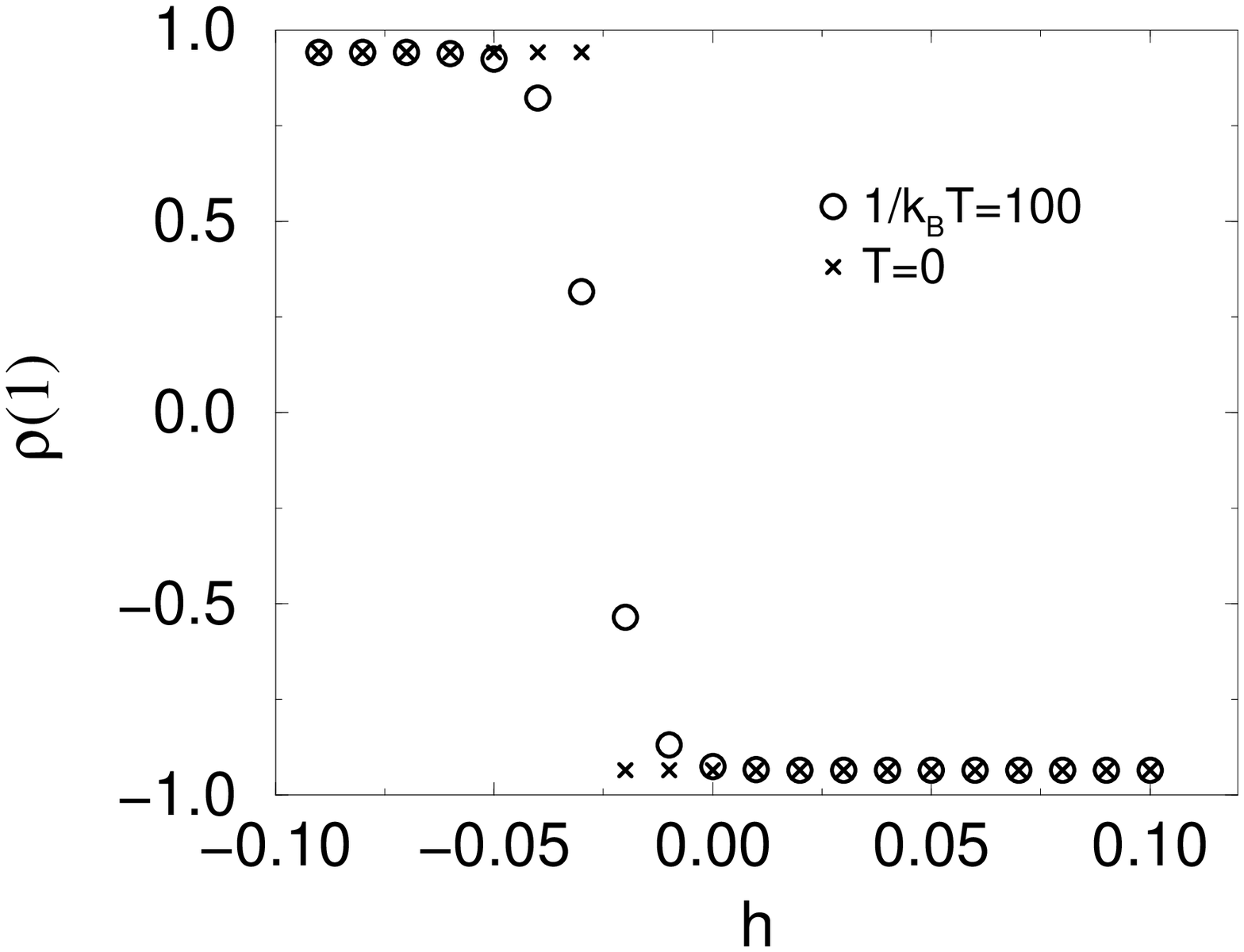}
 \caption{}\label{fig1}
\end{figure}

\begin{figure}
  \includegraphics[width=5.0in]{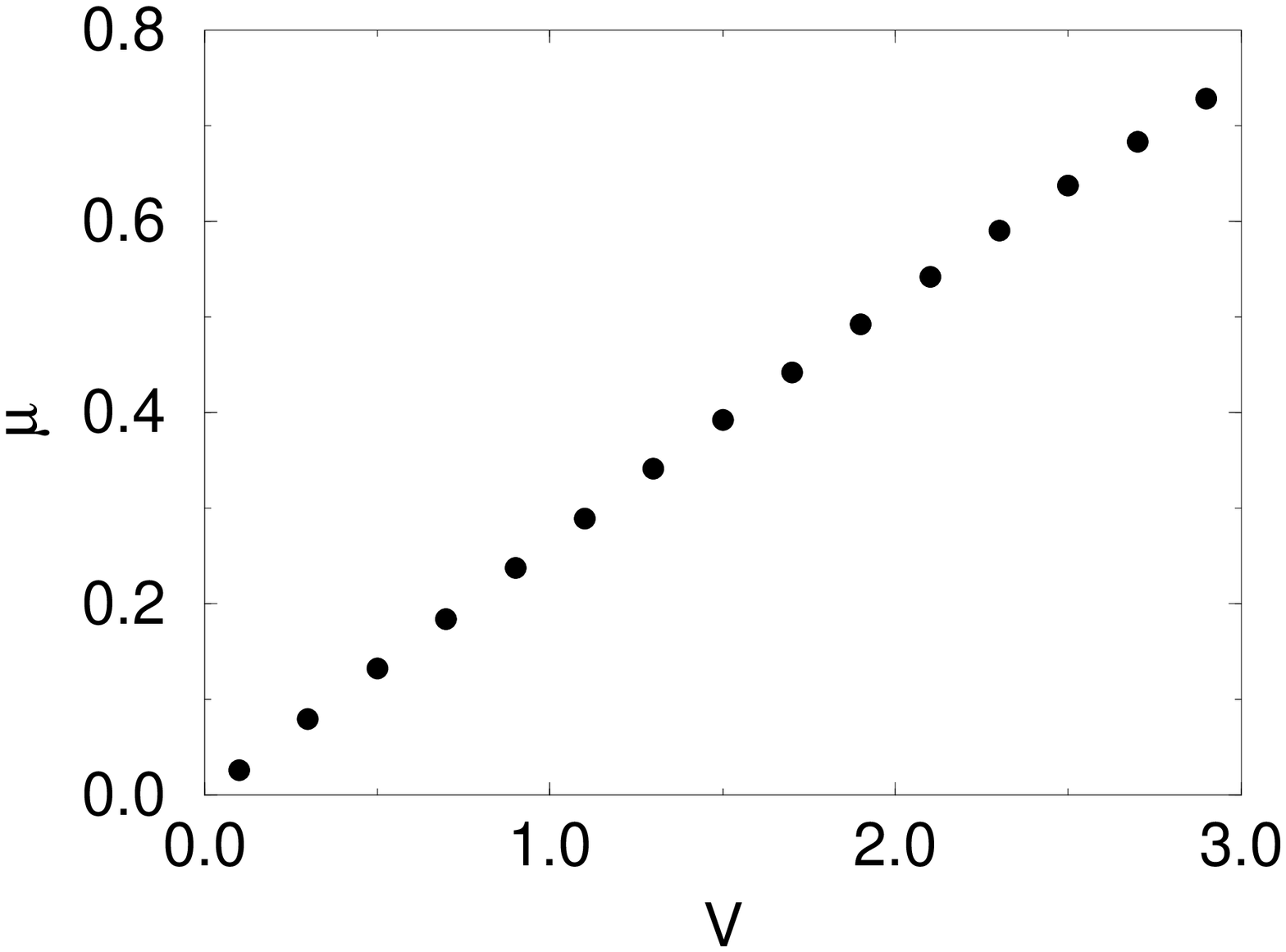}
\caption{}\label{fig2}
\end{figure}


\begin{thebibliography}{9}
\bibitem{uniform} In the case when $t_1 = t_2 $, our convention of the
Fourier transfomation is $\psi(x) = e^{i k_f x} \psi_R (x) +
 e^{-i k_f x} \psi_L (x)$, and the kinetic term is
$S_0 = \sum_{k,\omega}  {\psi_R}^{\dagger} (i \omega +v_f k)
{\psi_R} + {\psi_L}^{\dagger} (i \omega - v_f k)
{\psi_L}$.
\end{thebibliography}
\end{document}